\documentclass[aps,pre,twocolumn,floatfix]{revtex4-2}

\usepackage{epsf}
\usepackage{subfigure}
\usepackage{amsmath}
\usepackage{amssymb}
\usepackage{bm}
\usepackage{graphicx}
\usepackage{epstopdf}
\newcommand{\be}{\begin{equation}}
\newcommand{\ee}{\end{equation}}
\newcommand{\bea}{\begin{eqnarray}}
\newcommand{\eea}{\end{eqnarray}}

\newcommand{\parent}[1]{\left( #1 \right)}

\begin{document}

\title{\bf Irreversibility as divergence from equilibrium}

\author{David Andrieux}

\begin{abstract}
The entropy production is commonly interpreted as measuring the distance from equilibrium. 
However, this explanation lacks a rigorous description due to the absence of a natural equilibrium measure. 
The present analysis formalizes this interpretation by expressing the entropy production of a Markov system as a divergence with respect to particular equilibrium dynamics.
These equilibrium dynamics correspond to the closest reversible systems in the information-theoretic sense. 
This result yields novel links between nonequilibrium thermodynamics and information geometry.
\end{abstract}

\maketitle

\vskip 0,25 cm

\section{Context and objectives}

The study of nonequilibrium systems is a cornerstone of statistical physics, with entropy production serving as a fundamental quantity to quantify irreversibility and dissipation. Traditionally, entropy production is formulated in terms of thermodynamic fluxes and their conjugate affinities, providing a macroscopic framework to describe energy dissipation and transport phenomena in systems driven by external forces \cite{NP77, GM11}.   

At the mesoscopic scale, a deeper connection emerges between entropy production and the underlying microscopic dynamics. Here, the entropy production can be expressed through the transition rates and state probabilities governing the system's stochastic evolution \cite{S76, JVN84, H05}. In particular, this formulation links the entropy production with the breaking of time-reversal symmetry: the asymmetry in the likelihood of observing a trajectory versus its time-reversed counterpart is directly quantified by the entropy production along that trajectory \cite{G04, A07, A08}. This insight not only bridges thermodynamics with stochastic dynamics but also establishes entropy production as a measure of temporal irreversibility at the microscopic level. Similarly, mesoscopic systems subjected to time-dependent driving forces exhibit universal relations that constrain the statistics of work and other thermodynamics observables \cite{J97, C99}. 
These relations also arise from a comparison between forward trajectories and their time-reversed counterparts (under a time-reversed protocol), underscoring the central role of time-reversal symmetry breaking in nonequilibrium thermodynamics. 

At the same time, entropy production is often intuitively interpreted as a measure of "distance" from equilibrium. However, this interpretation lacks a rigorous formalization and faces conceptual challenges. The primary difficulty arises from the absence of a well-defined reference equilibrium from which to define such a distance. 
This non-uniqueness renders the notion of entropy production as a distance ambiguous and context-dependent. 
Consequently, while entropy production robustly quantifies irreversibility, its interpretation as a geometric distance remains elusive.

In this communication, I express the steady state entropy production as the symmetrized Kullback-Leibler divergence between the actual dynamics and two reference equilibrium dynamics. 
This finding provides a rigorous basis for the interpretation of the entropy production as measuring a distance from equilibrium, and deepens the connection between information geometry and nonequilibrium dynamics.

\section{Time-reversal and entropy production of Markov chains}

Let's consider a Markov chain characterized by a transition matrix $P = (P_{ij})$ on a finite state space of size $N$ (our results directly extends to continuous time processes).
The Markov chain is primitive, i.e., $P^{n}$ has all positive entries for $n$ larger than some $n_0$.
The chain $P$ then admits a unique stationary distribution $\pi$.

The entropy production of $P$ in the steady state $\pi$ takes the form \cite{NP77, S76, H05}
\bea
\Delta_i S
= \sum_{ij} \pi_i P_{ij} \ln \frac{P_{ij}}{P_{ji}} \geq 0 \, .
\label{EP}
\eea
This expression only depends on $P$ and its stationary distribution $\pi$. 
It vanishes for a reversible chain where detailed balance is satisfied, $\pi_i P_{ij} = \pi_j P_{ji}$ for all $i,j$.

The entropy production (\ref{EP}) involves the reversed probabilities $P_{ji}$, which are proportional to the time-reversed dynamics
\bea
P_{ij}^* = \frac{\pi_j}{\pi_i} P_{ji} \, .
\nonumber
\eea
The time-reversed dynamics has the same stationary distribution, i.e. $\pi^* = \pi$.
Intuitively, irreversibility thus arises from the asymmetry between the forward and the time-reversed processes.
This is formalized by writing the entropy production as \cite{G04, A07}
\bea
\Delta_i S = D(P|P^*) \, ,
\label{EP.HR}
\eea
where
\bea
D(P|Q) = \sum_{ij} \pi_i P_{ij} \ln \frac{P_{ij}}{Q_{ij}} \geq 0
\nonumber
\eea
is the Kullback-Leibler (KL) divergence rate between $P$ and $Q$.
Here the KL divergence is extended to first-order Markov chains, which can be shown to be well-defined, non-negative, and vanishing only when $P=Q$ \cite{MS94, S96}.
Note that in this case $D(P|P^*) = D(P^*|P)$ even though $D$ is not symmetric in general.

\section{Irreversibility as divergence from equilibrium}

I now introduce a novel formulation of the entropy production. 
In this formulation, the entropy production measures the dissimilarity between $P$ and specific equilibrium dynamics: 
\bea
\frac{1}{2}\Delta_i S = D_S\parent{P,P^{x}} \, ,
\label{EP.new}
\eea
where 
\bea
D_S(P, Q) = D(P|Q) + D(Q|P)
\nonumber
\eea
is the symmetrized KL divergence \cite{FN02}.
The relationship (\ref{EP.new}) holds for the two equilibrium dynamics \cite{FN03}
\bea
P^{m} = (P+P^*)/2 \quad {\rm and} \quad P^{e} = s\Big[ \parent{ P \circ P^{*} }^{(1/2)} \Big]  \, .
\nonumber
\eea
Here $\circ$ denotes the Hadamard product, $P^{(1/2)}$ is the elementwise exponentiation, and the mapping $s$ transforms a positive matrix $Q$ into a stochastic one as $s[Q] = (1/\rho)\, {\rm diag}(\alpha)^{-1} \, Q \, {\rm diag}(\alpha)$, where $\rho$ is the largest eigenvalue of $Q$ and $\alpha$ its corresponding right eigenvector.
Expression (\ref{EP.new}) is our central result.

The two dynamics $P^{m}$ and $P^{e}$ play a special role in information geometry, as highlighted in ref. \cite{WW21} for the standard KL divergence. 
Indeed, $P^{m}$ and $P^{e}$ correspond to the closest equilibrium systems to $P$ in the information-theoretic sense \cite{A16, WW21}:
\bea
P^{m} = \arg\min_{Q} D(P|Q),  \quad  P^{e} = \arg\min_Q D(Q|P) \, ,
\nonumber
\eea
where the minimization is performed over the space $\Sigma$ of compatible equilibrium dynamics \cite{FN01}. 

Here, the {\it symmetrized} divergence $D_S$ connects these concepts to nonequilibrium thermodynamics. 
This connection is further discussed in the last section while Eq. (\ref{EP.new}) is demonstrated in the appendix.




\section{A molecular motor model example}

Consider a Markov chain representing a molecular motor with $2\ell$ states corresponding to different conformations of the protein complex. 
These states form a cycle of periodicity $2\ell$ corresponding to a revolution by 360° for a rotary motor or a reinitialization step for a linear motor.
The motor alternates between two types of states according to the transition matrix \cite{AG06}
\bea
P=
\begin{pmatrix}
0 &  p_1 &  &  &  & 1-p_1\\
1-p_2 &  & p_2 &  &  & \\
 &  1-p_1 &  & p_1 & & \\
 &   & \ddots & 0 & \ddots & \\
 &   &  & 1-p_1 & 0 & p_1\\
p_2 &   &  &  & 1-p_2 & 0
\end{pmatrix}_{2\ell \times 2\ell}
\nonumber
\eea
The matrix $P$ is doubly stochastic, so that its stationary state the uniform distribution ${\bf \pi}= (1,1,\cdots ,1)/2\ell$ for all parameters $p_1,p_2$.

Given the symmetries of the model, a dynamics $P$ is determined by the two parameters $(p_1, p_2)$.
The time-reversed chain $P^* = (1-p_2, 1-p_1)$ since the steady state is uniform.
The equilibrium reference dynamics $P^x$ can also be determined analytically: 
\bea
P^m = \frac{1}{2}(p_1-p_2+1, p_2-p_1+1)
\nonumber
\eea
and 
\bea
P^e = \frac{1}{\Sigma} (\sqrt{p_1 (1-p_2)} , \sqrt{p_2 (1-p_1)}) \, ,
\nonumber
\eea 
where $\Sigma = \sqrt{p_1 (1-p_2)} + \sqrt{p_2 (1-p_1)}$. 

A direct calculation then shows that
\bea
D_S(P^m, P) = D_S(P^e, P) = (1/2)\Delta_i S \, ,
\nonumber
\eea 
with the entropy production 
\bea
\Delta_i S = \frac{1}{2} (p_1+p_2-1) \ln \parent{ \frac{p_1p_2}{(1-p_1)(1-p_2)} } \, .
\nonumber
\eea
As expected, the entropy production is given by the product $J \times A$, with $J = (p_1+p_2-1)/2\ell$ the average current and $A = \ell \ln [p_1p_2/((1-p_1)(1-p_2))]$ the corresponding affinity \cite{S76, AG06}.


\section{Perspective: Nonequilibrium transport and information geometry} 

\begin{figure}[t]
\includegraphics[scale=.42]{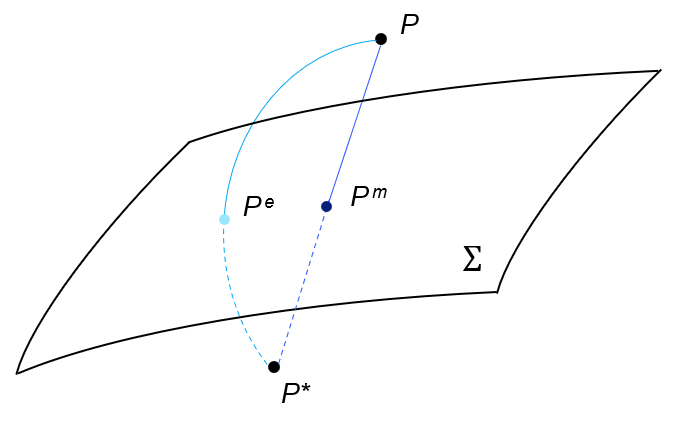}
\caption{{\bf Geometry of the space of Markov chains}. 
The set of equilibrium dynamics is represented as a two-dimensional manifold $\Sigma$. 
The entropy production is given by the symmetrized KL divergence $D_S$ between $P$ and $P^{e}$ or $P^{m}$, and is obtained by integrating the Fisher information along the $e$-geodesic and the $m$-geodesic (solid lines).
By symmetry, the integration can also be performed along the geodesics connecting $P^{e}$ or $P^{m}$ to $P^*$ (dashed lines), leading to the alternative formula $\Delta_i S = 2 D_S(P^x, P^*)$. 
The KL divergences between $P, P^m,$ and $P^e$ satisfy various inequalities, providing lower bounds for the entropy production (Eqs. (\ref{ineq1})-(\ref{ineq2}) in appendix).
}
\label{fig1}
\end{figure}

The result (\ref{EP.new}) establishes that irreversibility can be interpreted as an 'information divergence' with respect to equilibrium.
The reference equilibrium dynamics $P^{m}$ and $P^{e}$, identified as the closest reversible systems under the standard KL divergence, naturally reside on the $m$ and $e$ geodesics central to information geometry  (Figure~\ref{fig1}) \cite{N05, A16, WW21}. 

However, the connection between these reference dynamics and entropy production only emerges when we consider the symmetrized KL divergence. This symmetrized divergence provides a natural geometric structure for exponential probability distributions \cite{Tal20} and, similarly, gives rise to the dynamics $P^e$ and $P^m$ along with their corresponding geodesics within the space of Markov dynamics \cite{A24c}. Remarkably, the resulting geometric framework yields fully symmetric transport properties, even for systems driven far from equilibrium \cite{A12, A22}.

While this rich interplay between entropy production, transport and information geometry holds for stationary states, the role of the symmetrized divergence and of the equilibrium dynamics $P^{e}$ and $P^{m}$ in time-driven processes remains an open question.
In that direction, we note that the symmetrized KL divergence is obtained by integrating the Fisher information along the $e$ and $m$ geodesics (Theorem 3.2 in \cite{A16}). This is particularly significant given the central role of Fisher information in the thermodynamic control of systems operating in the linear regime \cite{Zal12}. This link suggests a concrete pathway for extending our geometric framework to encompass time-dependent, nonlinear processes.


Taken together, these findings deepen the connection between thermodynamics and information geometry, and point to novel geometric structures impacting transport phenomena, especially far from equilibrium \cite{A24c, A25}. \\

\vskip 0.2 cm

\noindent {\bf Data availability}. This work proceeds within a theoretical and mathematical approach and does not analyse or generate any datasets.

\vskip 0.2 cm

\noindent {\bf Conflict of interest}. The author declares that there are no conflicts of interest regarding the publication of this paper.



\section*{Appendix: Demonstration of the main result (\ref{EP.new})}


Using the Pythagorean identities
\bea
D(P|Q) = D(P|P^{m}) + D(P^{m}|Q)
\nonumber
\eea 
and
\bea
D(Q|P) = D(Q|P^{e}) + D(P^{e}|P) \, ,
\nonumber
\eea
each valid for a reversible $Q$ (Theorem 6.1 in ref. \cite{WW21}), we obtain that
\bea
D (P^{e}|P ) + D (P|P^{e} ) = D (P^{m}|P ) + D (P|P^{m} )
\nonumber
\eea
or 
\bea
D_S(P^{e}, P) = D_S(P^{m}, P) \, .
\nonumber
\eea
That is, if the relationship (\ref{EP.new}) holds for either $P^{m}$ or $P^{e}$, the other follows automatically. 
Nonetheless, it is instructive to prove both cases explicitly.
To this end, we introduce the relative entropies
\bea
h(P|Q) = - \sum_i \pi_i P_{ij} \ln Q_{ij}
\nonumber
\eea
so that
\bea
D(P|Q) = h(P|Q) - h(P|P) \, .
\nonumber
\eea

Let's first demonstrate the case $P^{x} = P^{e}$.
The symmetrized divergence reads
\bea
D_S(P,P^{e}) = D (P^{e}|P ) + D (P|P^{e} ) \, .
\nonumber
\eea
The same log ratios $\ln (P_{ij}/P^{e}_{ij})$ appear in both terms on the right hand side, and take the form
\bea
\ln \frac{P_{ij}}{P^{e}_{ij}} = \frac{1}{2} \ln \frac{P_{ij}}{P_{ji}} +\frac{1}{2} \ln \frac{\pi_i}{\pi_j} + \ln \frac{\alpha_i}{\alpha_j} + \ln \rho \, , \nonumber
\eea
where we used that $P^{e}_{ij} = (1/\rho)\, (\alpha_j/\alpha_i) \sqrt{P_{ij}P_{ji} (\pi_j/\pi_i)}$.
Inserting this expression into $D (P^{e}|P )$ we get that
\bea
D (P^{e}|P ) &=& (1/2) [h(P^{e}|P) - h(P^{e}|P^*)] - \ln \rho \nonumber \\
&=& - \ln \rho \, .\nonumber 
\eea
Here we used that the terms $\ln (\pi_i/\pi_j)$ and $\ln (\alpha_i/\alpha_j)$ vanish when averaged over a stochastic dynamics (see for example Lemma 4.3 (iii) in ref. \cite{WW21}) to get the first equality. 
For the second equality, Lemma 4.3 (ii) from reference \cite{WW21} shows that $h(P^{e}|P) - h(P^{e}|P^*) = 0$ since $P^{e}$ is reversible and the log ratios $\ln (P_{ij}/P_{ji})$ are antisymmetric in $(i,j)$.

In parallel we have
\bea
D (P|P^{e} ) &=& (1/2)[h(P|P^*) - h(P|P)] + \ln \rho \nonumber \\
&=& (1/2) \Delta_i S + \ln \rho \, .\nonumber
\eea
Here we also used that the terms $\ln (\pi_i/\pi_j)$ and $\ln (\alpha_i/\alpha_j)$ vanish when averaged over a stochastic dynamics.
The last equality uses that $h(P|P^*) - h(P|P) = D(P|P^*)$ is the entropy production (\ref{EP}).
Summing the last two equations the terms $\pm \ln \rho$ cancel each other and we obtain Eq. (\ref{EP.new}).  $\square$\\

Let's now demonstrate Eq. (\ref{EP.new}) when $P^{x} = P^{m}$.
We have
\bea
D_S(P^{m}, P) &=& [h(P|P^{m}) - h(P|P)] \nonumber \\
& & + [h(P^{m}|P) - h(P^{m}|P^{m})] \nonumber \\
&=& (1/2) [h(P|P^*) - h(P|P)] \nonumber \\
& & + (1/2) [h(P|P^{m}) - h(P^*|P^{m})] \nonumber \\
&=& (1/2) [h(P|P^*) - h(P|P)] \nonumber \, .
\eea
The first equality is obtained by expressing the KL divergences in terms of relative entropies.
In the second equality, we used that $h(P^{m}|Q) = (1/2)h(P|Q) + (1/2)h(P^*|Q)$ and that $h(P^*|P) = h (P|P^*)$. These two relations hold because $\pi$ is also the stationary distribution of $P^*$ and $P^{m}$.
The third equality comes from Lemma 4.3 (i) in reference \cite{WW21}.
Then, the last expression $h(P|P^*) - h(P|P) = D(P|P^*)$ is the entropy production (\ref{EP.HR}). $\square$ \\

We also note that the Pythagorean equalities, together with the positivity of the KL divergence, constrain the different geometric components and bound the entropy production from below: 
\bea
\Delta_i S &\geq& 2 D \parent{P^{m}|P} \nonumber \\
&\geq& 2 \max \left[ D \parent{P^{e}|P}, D \parent{P^{m}|P^{e}} \right] \nonumber \\
&\geq& 0
\label{ineq1}
\eea
and
\bea
\Delta_i S &\geq& 2 D \parent{P|P^{e}} \nonumber \\
&\geq& 2 \max \left[ D \parent{P|P^{m}}, D \parent{P^{m}|P^{e}} \right] \nonumber \\
&\geq& 0 \, .
\label{ineq2}
\eea



\begin{thebibliography}{99}
\bibitem{FN02} While $D_S(P,Q)$ is symmetric, non-negative, and vanishes when $P = Q$, it doesn't define a distance since the triangle inequality is not always respected.
\bibitem{FN03} The dynamics $P^m$ is an equilibrium dynamics since it satisfies the detailed balance condition with its steady state distribution $\pi^m = \pi$. To verify that the dynamics $P^e$ is an equilibrium dynamics, since $\pi^e \neq \pi$ in general, it is easier to check that it satisfies the equivalent Kolmogorov conditions, i.e. that the affinity of an arbitrary cycle vanishes.
\bibitem{FN01} A stochastic matrix $Q$ is said to be compatible with $P$ if it satisfies $Q_{ij} \geq 0$ if $P_{ij} \geq 0$ and $Q_{ij} = 0$ if $P_{ij} = 0$.
\end{thebibliography}
\end{document}